\documentclass[dvips,12pt]{article}
\usepackage{graphicx,color}
\newcommand\figcaption{\def\@captype{figure}\caption}
\newcommand\tabcaption{\def\@captype{table}\caption}
\def\pa{\partial}
\def\al{\alpha}

\def\Ga{\Gamma}

\def\la{\lambda}

\def\be{\beta}
\def\Dl{\Delta}
\def\kp{\kappa}
\def\th{\theta}

\def\vf{\varphi}
\def\om{\omega}
\def\nb{\nabla}
\def\veps{\varepsilon}

\def\nn{\nonumber}

\def\diag{\mbox {diag}}
\title{\Large\bf Exact Vacuum Solutions to the Einstein Equation}
\author{Ying-Qiu Gu\footnote {email: yqgu@luody.com.cn}}
\date{\small School of Mathematical Science, Fudan University, Shanghai 200433, China}

\begin{document}
\maketitle\DeclareGraphicsRule{.eps.gz}{eps}{.eps.bb}{`gunzip -c
#1}
\begin{abstract}
In this paper, we present a framework for getting a series of
exact vacuum solutions to the Einstein equation. This procedure of
resolution is based on a canonical form of the metric. According
to this procedure, the Einstein equation can be reduced to some
2-dimensional Laplace-like equations or rotation and divergence
equations, which are much convenient for the resolution.


\vskip 1.0cm \large {2000 MR Subject Classification:
83C05,~83C15,~83C57}
\end{abstract}

\section{Introduction}
\setcounter{equation}{0}

The Einstein equation
\begin{equation}
G_{\mu\nu}\stackrel{{\mathrm{def}}}{=} R_{\mu\nu}-\frac 1
2g_{\mu\nu} R =\kp T_{\mu\nu} \label{1.1}
\end{equation}
is a highly nonlinear system of partial differential equations
satisfied by the metric $g_{\mu\nu}$, where  $R_{\mu\nu}$ stands
for the Ricci tensor and $R$ the scalar curvature of the spacetime
manifold defined respectively by
\begin{equation}
R_{\mu\nu}=\pa_\al\Ga^\al_{\mu\nu}-\pa_\mu\Ga^\al_{\nu\al}+
\Ga^\al_{\mu\nu}\Ga^\be_{\al\be}-\Ga^\al_{\mu\be}\Ga^\be_{\nu\al},
\label{1.2}
\end{equation}
and
\begin{eqnarray}
R= g^{\mu\nu}R_{\mu\nu}, \label{1.3}
\end{eqnarray}
in which $\Ga^\al_{\mu\nu}$ is the Christoffel symbol given by
\begin{equation}
\Ga^\al_{\mu\nu}=\frac 1 2 g^{\al\be}(\pa_\mu g_{\be\nu}+\pa_\nu
g_{\mu\be}-\pa_\be g_{\mu\nu}), \label{1.4}
\end{equation}
$T_{\mu\nu}$ is the energy-momentum tensor of matter, and the
Greek indexes go from 0 to 3 or from $t, z, x$ to $y$. In the
vacuum domain we have
\begin{equation}
T_{\mu\nu}\equiv 0.\end{equation}

To get the exact vacuum solution of the Einstein equation is an
arduous work\cite{1,2}. The conventional method of resolution is
to analyze the symmetry of the metric, i.e. analyze the Killing
vectors of the spacetime. Almost all the well-known solutions such
as the Friedmann-Robertson-Walker metric, Bianchi universe,
Lemaitre-de Sitter universe, as well as Schwarzschild metric and
Kerr metric, Taub-NUT solution\cite{2,3,5}, are based on the
symmetry of spacetime.

Obviously, a good choice for the coordinate system of the
spacetime manifold should be very helpful for analyzing the
Einstein equation. The traditional choices are the Gaussian
coordinate system or the harmonic coordinate system\cite{6}. The
former gives locally the following metric
\begin{equation} g_{\mu\nu}= \left( \begin {array}{cccc}1&0&0&0
\\\noalign{\medskip} 0&g_{{1 1}}&g_{{1 2}}&g_{{1 3}}
\\\noalign{\medskip}0&g_{{2 1}}&g_{{2 2}}&g_{{2 3}}
\\\noalign{\medskip}0&g_{{3 1}}&g_{{3 2}}&g_{{3 3}}
\end {array} \right),
\label{gauss} \end{equation} while, the latter satisfies the
following coordinate condition
\begin{equation}
g^{\al\be}\Ga^\mu_{\al\be}=0.\label{harm}\end{equation} However
(\ref{gauss}) and (\ref{harm}) are more convenient only for some
theoretical analysis rather than for practical resolution.

Fortunately, the problem can be much simplified in a specially
chosen coordinate system. Under some natural assumptions we have
found that the metric can be transformed into the following
canonical form
\begin{equation} g_{\mu\nu}=\left( \begin {array}{cccc} {u}&v&p&q\\ \noalign{\medskip}v&0&0&0\\
\noalign{\medskip}p&0&-{a}&0\\ \noalign{\medskip}q&0&0&-{b}
\end {array} \right)
\label{1.01} \end{equation} with
\begin{equation}\det(g_{\mu\nu})=-v^2ab
\label{1.02} \end{equation} and the inverse
\begin{equation}
g^{\mu\nu}=\left( \begin {array}{cccc} 0&{v}^{-1}&0&0\\
\noalign{\medskip}{v}^{-1}&-{\frac
{uab+{p}^{2}b+{q}^{2}a}{{v}^{2}ab}}&{\frac {p}{va}}&{\frac {
q}{vb}}\\\noalign{\medskip}0&{\frac {p}{va}}&-{a}^{-1}&0
\\\noalign{\medskip}0&{\frac {q}{vb}}&0&-{b}^{-1}\end {array}
\right), \label{1.03} \end{equation} where $u, v, p, q, a, b$ are
smooth functions of the coordinates $(t,z,x,y)$ (See \cite{prf}
for the proof of this canonical form).

The canonical form (\ref{1.01}) is an Alibaba's conjuration for
solving the Einstein equation. In what follows, we will show how
to use this metric to solve the vacuum Einstein equation.

\section{Procedure of Resolution}
\setcounter{equation}{0} For the canonical metric (\ref{1.01}),
computing the Einstein tensor $G_{\mu\nu}$, we get
\begin{equation}
G_{zz}=-\frac 1 2 \left[\left(\frac{\pa b }{b\pa z}+\frac{\pa
a}{a\pa z}\right)\frac{\pa v}{v\pa z} +\frac 1 2\left(\frac{\pa
a}{a\pa z}\right)^2+\frac1 2 \left(\frac{\pa b}{b\pa z}\right)^2
-\frac {\pa^2 a }{a\pa z^2}-\frac {\pa^2 b }{b\pa z^2}\right].
\label{2.2}  \end{equation} If $\pa_z(ab)\ne 0$, solving
$G_{zz}=0$ with respect to $v$, we get
\begin{equation}
v=V \exp\left[{\int}\left(\frac {\pa^2 a }{a\pa z^2}+\frac {\pa^2
b }{b\pa z^2} -\frac 1 2\left(\frac{\pa a}{a\pa z}\right)^2-\frac1
2 \left(\frac{\pa b}{b\pa z}\right)^2 \right)\left(\frac{\pa b
}{b\pa z}+\frac{\pa a}{a\pa z}\right)^{-1}dz\right], \label{2.3}
\end{equation} where $V=V(t,x,y)$ is a function to be determined.

In order that the following recursive procedure goes on, for the
exact   solutions, we must adopt an ansatz for $a$ and $b$ to make
(\ref{2.3}) integrable. Different ansatz leads to different family
of solutions. In what follows we take the separating variable
ansatz as a representative example to show how the procedure
works, that is, we assume
\begin{equation}
a=L(t,x,y) K(t,z)^2,\qquad b=N(t,x,y) K(t,z)^2,
\label{2.4}\end{equation}where $\pa_z K\ne 0$.

Substituting (\ref{2.4}) into (\ref{2.3}), it is easy to see that
\begin{equation}v=V\pa_z K.
\label{2.5}  \end{equation}

Using (\ref{2.4}) and (\ref{2.5}),  we have
\begin{equation}
G_{zx}=-\frac 1 {2V}\left(\frac{\pa^2_z p}{\pa_z K}-\frac{\pa^2_z
K \pa_z p}{(\pa_z K)^2} - 2p\frac{\pa_z K}{K^2} +2 \pa_x V\frac
{\pa_z K}{K}\right). \label{2.6}\end{equation} Solving $G_{zx}=0$
with respect to $p$, we get
\begin{equation}
p= P K^2 +\frac {\pa V}{\pa x}K+\frac A K, \label{2.7}
\end{equation} where $P, A$ are functions of $t, x, y$ to be determined.

Similarly, we have
\begin{equation}
G_{zy}=-\frac 1 {2V}\left(\frac{\pa^2_z q}{\pa_z K}-\frac{\pa^2_z
K \pa_z q}{(\pa_z K)^2} - 2q\frac{\pa_z K}{K^2} +2 \pa_y V\frac
{\pa_z K}{K}\right). \label{2.8}  \end{equation} Solving
$G_{zy}=0$ with respect to $q$, we get
\begin{equation}
q=- Q K^2 +\frac {\pa V}{\pa y}K+\frac B K, \label{2.9}
\end{equation} where $Q, B$ are functions of $t, x, y$ to be determined.

Substituting (\ref{2.4}), (\ref{2.5}), (\ref{2.7}) and (\ref{2.9})
into $G_{xy}=0$, we get
\begin{eqnarray}
\pa_x Q-\pa_y P =-P \frac {\pa_y L}L+Q\frac {\pa_x N}N
\label{2.10}\end{eqnarray} and
\begin{eqnarray}
A=B=0. \label{AB}  \end{eqnarray} By $G_{xx}=0$ or $G_{yy}=0$, we
obtain
\begin{equation}
\frac{\pa_x P} L+\frac{\pa_y Q} N-\frac P {2L} \left(\frac{\pa_x
L} L+\frac{\pa_x N} N\right)-\frac Q {2N} \left(\frac{\pa_y L}
L+\frac{\pa_y N} N\right)=-\frac 1 2 \left(\frac{\pa_t L}
L-\frac{\pa_t N} N\right). \label{2.11}  \end{equation} Therefore,
taking $(P,~Q)$ as a vector in subspace $(x, y)$, (\ref{2.10}) and
(\ref{2.11}) provide the corresponding covariant rotation and
divergence equations respectively.

By $G_{tz}=0$, we get
\begin{equation}
u=-\left(\frac {P^2} L+\frac {Q^2}
N\right)K^2+2\pa_t(VK)+WVK+U+\frac \al {VK}, \label{2.12}
\end{equation} where $\al$ is a function independent of $z$ to be determined.
$U(t,x,y)$ and $W(t,x,y)$ satisfy
\begin{equation}
\Dl V = \frac {U+|\nb V|^2} V-SV \label{2.13}
\end{equation}
with
\begin{equation}
S = \frac1 4 \left(\frac {(\pa_x N)^2}{LN^2}+\frac {(\pa_y
L)^2}{L^2N}+\frac {\pa_x L\pa_x N}{L^2N}+\frac {\pa_y L\pa_y
N}{LN^2}-\frac 2 {LN} (\pa^2_x N+\pa^2_y L)\right),\label{eqs}
\end{equation}
and
\begin{eqnarray}
\frac{\pa_x P} L-\frac{\pa_y Q} N &-& \frac P {2L}
\left(\frac{\pa_x L} L-\frac{\pa_x N} N\right)-\frac Q {2N}
\left(\frac{\pa_y L} L-\frac{\pa_y N} N\right)+\frac 1 2
\left(\frac{\pa_t L}
L+\frac{\pa_t N} N\right) \nn \\
 &=&W+\frac 2 V \left(\pa_t V+\frac 1 L
{P\pa_x V}-\frac 1 N {Q\pa_y V}\right), \label{2.14}
\end{eqnarray} in which the covariant Laplace and gradient operators of
a scalar in the subspace $(x, y)$ are defined respectively by
\begin{eqnarray} \Dl &=& \frac 1 L {\pa_x^2 }+\frac 1 N {\pa^2_y }+
\frac 1 2 \left(-\frac {\pa_x L} {L^2}+\frac {\pa_x N}
{LN}\right)\pa_x +\frac 1 2 \left(\frac {\pa_y L}{LN}-\frac {\pa_y
N} {N^2}\right)\pa_y,\\
\nb &=& \left(\frac 1{\sqrt L}\pa_x,~\frac 1{\sqrt N}\pa_y\right).
\label{2.15}  \end{eqnarray}

By $G_{tx}=G_{ty}=0$, we find $\al=\al(t)$. We should point out
that, $G_{tx}=G_{ty}=0$ also provide two linear Laplace-like
equations of $P$ and $Q$, which are consequences of (\ref{2.10})
and (\ref{2.11}).

For $G_{tt}=0$, a direct calculation leads to
\begin{eqnarray}
\Dl U &=& -\frac {3W\al+2\pa_t\al}{V^2}, \label{2.16} \\
\Dl W &=& -\frac 2 {V^2}\left( W U+\pa_t U+\frac 1 L P\pa_x
U-\frac 1 N Q\pa_y U \right). \label{2.17}  \end{eqnarray}

Thus, we can take $U,~V,~ W,~ P,~ Q,~\al$ as unknown functions
satisfying the almost linear partial differential equations
(\ref{2.16}), (\ref{2.13}), (\ref{2.17}), (\ref{2.10}),
(\ref{2.11}) and (\ref{2.14}), while, take $K(t,z),~L(t,x,y),
~N(t,x,y)$ as functions determined by coordinates, boundary
conditions and initial conditions. Any one of the equations
(\ref{2.16}), (\ref{2.13}) and (\ref{2.17}) is quite similar to
the Ernst equation\cite{Kln,ern}.

Furthermore, we can simplify the metric by setting  $r=K$ as new
coordinate to replace $z$ as follows. The line element of the
spacetime is then given by
\begin{eqnarray}
ds^2&=&udt^2+2dt(vdz+pdx+qdy)-(adx^2+bdy^2),\nn\\
&=&(u-2V\pa_t K) dt^2+2 V dt d K +2dt(pdx+qdy)-(adx^2+bdy^2),\nn\\
&=&\bar u dt^2+2 V dt dr+2dt(pdx+qdy)-(adx^2+bdy^2), \label{2.18}
\end{eqnarray} where
\begin{equation}
\bar u=-\left(\frac {P^2} L+\frac {Q^2} N\right)r^2+(2\pa_t
V+WV)r+U+\frac \al {Vr}. \label{2.19}
\end{equation}
Thus, in the new coordinate system $(t,r,x,y)$, the metric becomes
\begin{equation}
g_{\mu\nu}= \left( \begin {array}{cccc} \bar u & V  & P {r}^{2} +
\pa_x V r  & -Q {r}^{2} + \pa_y V r
\\ \noalign{\medskip} V & 0 & 0 & 0
\\ \noalign{\medskip} P {r}^{2}+ \pa_x V r & 0 & -L {r}^{2} & 0\\
\noalign{\medskip}- Q{r}^{2 }+ \pa_y V r & 0 & 0 & -N r^2 \end
{array} \right). \label{2.21} \end{equation}

Similarly, if we take $\rho= V K $ as new coordinate to replace
$z$, then in the new coordinate system $(t, \rho, x, y)$, the
metric becomes
\begin{equation}
g_{\mu\nu}= \left( \begin {array}{cccc} -\left(\frac {P^2} L+\frac
{Q^2} N\right)\frac {\rho^2} {V^2}+W\rho+U+\frac \al {\rho} & 1 &
\frac P {V^2} {\rho^2}  & -\frac Q {V^2}{\rho^2}
\\ \noalign{\medskip} 1 & 0 & 0 & 0
\\ \noalign{\medskip} \frac P{V^2} {\rho}^{2} & 0 & -\frac L {V^2} {\rho}^{2} & 0\\
\noalign{\medskip}- \frac Q {V^2} {\rho}^{2 } & 0 & 0 & -\frac N
{V^2} \rho^2
\end {array} \right).\label{2.22}
\end{equation}

The above procedure can provide a family of solutions depending on
4 coordinates, which is relatively complicated. However, for some
special cases including more symmetries, the procedure can be much
simpler. For instance, if we consider the static case with the
following ansatz
\begin{equation} a=b=a(x,y),\quad p=q=0,\quad
u=u(x,y), \quad v=v(x,y),\label{2.23} \end{equation} then the
solutions can be explicitly obtained. In fact, for this case,
$G_{tx},~G_{ty},~G_{zz},~G_{zx}$ and $G_{zy}$ vanish
automatically. By $G_{xx}+G_{yy}=0$, we get
\begin{equation}
(\pa_x^2+\pa_y^2)v=0,\label{2.24}
\end{equation}
that is to say, $v(x,y)$ is a 2-dimensional harmonic function,
which can be easily generated by means of complex analytic
function $f(x+yi)$.

Moreover, by $G_{xx}-G_{yy}=G_{xy}=G_{tt}=0$ and (\ref{2.24}), we
get
\begin{eqnarray}
(\pa_x^2+\pa_y^2)a &=& \frac 1 a(|\pa_x a|^2+|\pa_y a|^2)+\frac a
{2v^2}(|\pa_x v|^2+|\pa_y v|^2),\label{2.26} \\
(\pa_x^2+\pa_y^2)u
&=& \frac 1 v(\pa_x u\pa_x v+\pa_y u\pa_y v)-\frac u {v^2}(|\pa_x
v|^2+|\pa_y v|^2).\label{2.25}
\end{eqnarray}
The solution of (\ref{2.26}) consistent with
$G_{xx}=G_{xy}=G_{yy}=0$ is given by
\begin{equation}
a=\frac \la {\sqrt{v}}(|\pa_x v|^2+|\pa_y v|^2),\label{2.27}
\end{equation}
where $\la$ is a constant.

(\ref{2.25}) is a linear equation of $u$. Let $u=vw$, we get
\begin{eqnarray} (\pa_x^2+\pa_y^2)w =- \frac 1 v(\pa_x w\pa_x
v+\pa_y w\pa_y v).\label{2.25w}
\end{eqnarray}
For any given $v$, the solution of (\ref{2.25w}) can be easily
determined by an appropriate boundary condition. Here we provide
the analytic solution related to $v$, which is given by
\begin{equation}
u=v[m+n\ln v+(k+j\ln v)\bar v], \label{2.28}\end{equation} where
$j,k,m,n$ are constants, $\bar v$ is the conjugate harmonic
function of $v$, namely,
\begin{eqnarray}
v=\Re(f(x+yi)),&~~& \bar v=\Im(f(x+yi)),\end{eqnarray} or
\begin{eqnarray}
v=\Im(f(x+yi)),&~~& \bar v=\Re(f(x+yi)).\end{eqnarray}

For other ansatz of the metric, the recursive resolution procedure
is basically along the same line, that is
\begin{equation}
G_{zz} \to (G_{zx},~ G_{zy}) \to (G_{xy},~ G_{xx},~ G_{yy},~
G_{tz}) \to (G_{tx},~ G_{ty}) \to G_{tt}.\end{equation}

\section{Examples}
\setcounter{equation}{0}

{\bf Example 1.~~} For the case (\ref{2.23})$\sim$(\ref{2.28}), if
we take the complex function as
\begin{eqnarray}
f= \frac Z R-\frac R Z,\label{3.1}
\end{eqnarray}
where $Z = x+yi$ and $R$ is a real number, we have
\begin{eqnarray}
v=\Re(f)=\frac {x(x^2+y^2-R^2)}{R(x^2+y^2)},&~& \bar
v=\Im(f)=\frac
{y(x^2+y^2+R^2)}{R(x^2+y^2)}\label{3.2}\end{eqnarray} Substituting
(\ref{3.2}) into (\ref{2.27}) and (\ref{2.28}), we get
\begin{eqnarray}
a&=&\frac{\la(x^4+2x^2y^2+2R^2x^2+y^4+R^4-2R^2y^2)}{R\sqrt{x(x^2+y^2-R^2)(x^2+y^2)^3R}},\\
u&=&\frac{x(x^2+y^2-R^2)}{R(x^2+y^2)} \left(m+n\ln v+(k+j\ln
v)\frac{y(x^2+y^2+R^2)}{R(x^2+y^2)}\right).\end{eqnarray} This is
a more complicated version than that presented in \cite{gu}.

{\bf Example 2.~~} For the  case (\ref{2.23})$\sim$(\ref{2.28}),
if we take the complex function as
\begin{eqnarray}
f= \cosh\left(\frac {x+yi}R\right),
\end{eqnarray}
where $R$ is a real number, then we have
\begin{eqnarray}
v = \Re(f)=\cosh\frac x R \cos\frac y R, &\quad& \bar
v=\Im(f)=\sinh\frac x R\sin\frac y R.\label{3.5}
\end{eqnarray}
Substituting (\ref{3.5}) into (\ref{2.27}) and (\ref{2.28}), we
get
\begin{eqnarray}
a& =&\frac \la {R^2}\left(\cosh^2\frac x R-\cos^2\frac y
R\right)\sqrt{\left(\cosh\frac x R \cos\frac y R\right)^{-1}}, \\
u&=&\cosh\frac x R\cos\frac y R\left(m+n\ln v+(k+j\ln
v)\left(\sinh\frac x R\sin\frac y R\right)\right).\end{eqnarray}
Obviously, we can construct infinite solutions in this family by
choosing different analytic function $f$.

{\bf Example 3.~~} For the static case of metric (\ref{2.21}), if
we take the following ansatz
\begin{eqnarray}
L = 1,\qquad N=\sin^2 x,\qquad U=0,\label{3.9}
\end{eqnarray}
then $S=1$ in (\ref{2.13}) and we can get
\begin{eqnarray}
W=0,~~ P=m\sin x ,~~Q=k\sin^2 x,~~V=n\sin x \tan^\be \frac x 2 ,
\label{3.10} \end{eqnarray} where $k,~m,~n,~\be$ are constants.

If we take $k=\al, m=\be=0, n=1$ in (\ref{3.10}), under a suitable
coordinate transformation, from (\ref{3.9}) and (\ref{3.10}) we
can get a clear and normal form of metric (\ref{2.21}) as follows
\begin{equation} g_{\mu\nu}=\left( \begin {array}{cccc} {\frac \al
{r \sin x }}&  \sin x  &  r \cos x &0
\\\noalign{\medskip}  \sin x  &-{\frac {r
}{r-R} \sin^2 x }&0&{\sqrt{\frac {r }{r-R}}}r \sin^2 x
\\\noalign{\medskip}  r \cos x &0&-{
r}^{2}&0\\\noalign{\medskip}0&{\sqrt{\frac {r }{r-R}}} r\sin^2
x&0&-{r}^{2}
 \sin^2 x \end {array} \right),\label{1}
 \end{equation}
where $\al,~R$ are constants. In this case, the corresponding
spacetime has a torus structure.

{\bf Example 4.~~} For the dynamic case of metric (\ref{2.21}), if
we take the following ansatz
\begin{eqnarray}
L = 1,\quad N=\sin^2 x,\quad P=Q=\pa_y V=0,
\end{eqnarray}
then we can get
\begin{eqnarray}
W&=&-\frac 2 f f'(t),~~~U=kf^2,~~~\al = m f^3,\\
V&=&\frac {f \sin x}{2\be \sqrt{\be n}} \left(\be \tan^{-\be
}\frac x2 +kn\tan^\be \frac x 2\right),\end{eqnarray} or
\begin{eqnarray}
 V&=&\frac {f \sin  x}{2\be
\sqrt{\be n}} \left(n\be \tan^{-\be}\frac x2 +k\tan^\be \frac x
2\right),\label{3.15}\end{eqnarray} where $k,~m$ and $n\be>0$ are
constants, and $f(t)$ is an arbitrary function of $t$.

In particular, setting $k=1,\be=\frac 1 2, n=2$ and
\begin{equation}
f=1+\veps\sin(\om t), \end{equation} using (\ref{3.15}) we get
\begin{equation} V=(1+\veps\sin(\om t))\sin x
\left(\sqrt{\tan\frac x 2}+\sqrt{\tan^{-1} \frac x 2}\right),
\end{equation} then metric (\ref{2.22}) becomes
\begin{equation}
g_{\mu\nu}= \diag \left\{\left( \begin {array}{cc} f^2+\frac {m }
{\rho}f^3-\frac 2 f{f'}\rho & 1
\\ \noalign{\medskip} 1 & 0\end {array} \right),~-\frac  {\rho^2} {V^2} ,~ -\frac {\rho^2} {V^2}
\sin^2 x \right\} ,\label{3.16}
\end{equation}
where $\veps$ is a constant, $(t,\rho, x, y)$ are the coordinates.
This metric provides a rigorous model for the study of the
behavior of gravitational waves.

{\bf Example 5.~~}  For the periodic case of metric (\ref{2.21}),
if we take the following ansatz
\begin{eqnarray}
L = 1,\qquad N=\sin^2 x,\qquad V=1,
\end{eqnarray}
then we can get
\begin{eqnarray}
U=1,\qquad W&=&0,\qquad \al =-2m,\\
P=f \cos(y-h),&& Q=f \cos x \sin x\sin(y-h),\end{eqnarray} where
$m$ is a constant, and $f(t),~h(t)$ are arbitrary functions of
$t$. This metric provides a strange rotary gravity.

Setting $f=0$ and making a coordinate transformation
$$t=\tau-r-2m\ln(r-2m), ~~ x=\th, ~~ y=\vf,$$ metric (\ref{2.21}) becomes the
normal Schwarzschild metric in system $(\tau, r, \th, \vf)$:
\begin{equation}
g_{\mu\nu}=\diag \left\{ 1 -\frac {2m} r,~ -\left(1 -\frac {2m}
r\right)^{-1},~ r^2, ~ r^2\sin^2\th\right\}.
\end{equation}

{\bf Example 6.~~}  For metric (\ref{2.21}), if we take $L=N=1$,
then we have the following solutions with axial symmetry
\begin{eqnarray}
g_{tt}&=&\left( k- \frac {4\beta m} {\rho   r } \left( k{\rho
}^{\beta}+{\rho  }^{-\beta} \right)^{-1} \right) {f}^{2}
,\\
g_{tr }&=&\frac {\rho   f}{2\be} {  \left( k{\rho  }^{\beta}+{\rho  }^{-\beta} \right)},\\
g_{tx}&=&\frac {x r  f}{2\rho  \beta} \left[  \left( k{\rho
}^{\beta}-{\rho  }^{-\beta} \right) \beta+k{\rho }^{\beta}+{\rho
}^{-\beta}
\right],\\
g_{ty}&=&\frac {y r  f}{2\rho  \beta} \left[  \left( k{\rho
}^{\beta}-{\rho  }^{-\beta} \right) \beta+k{\rho }^{\beta}+{\rho
}^{-\beta} \right]
,\\
 g_{xx}&=&g_{yy}=-r ^2
\end{eqnarray}
or
\begin{eqnarray}
g_{\mu\nu}= \left( \begin {array}{cccc} -k^{2}r^{2}f^{2}+\frac 2 f
{ f'(t)  r}-\frac 2 r { m f^{-3}}&1&-k y {{r}^{2}{f^{2}}}&-k x
{{r}^{2}{f^{2}}}
\\\noalign{\medskip}1&0&0&0\\\noalign{\medskip}-{k y {{r}^{2}}{
 f^{2}}}&0&{ -\rho^2
{r}^{2}}f^{2}&0\\\noalign{\medskip}-k x
{r}^{2}{f^{2}}&0&0&{-\rho^2 {r}^{2}}{f^{2}}
\end {array} \right),
\end{eqnarray}
where $k,~m,~\be$ are constants, $\rho=\sqrt{x^2+y^2}$,  $f(t)$ is
an arbitrary function, and $(t,~r,~x,~y)$ are the coordinates.

For cases without axial symmetry, we can get solutions as follows:
\begin{eqnarray}
g_{\mu\nu}=\left( \begin {array}{cccc} 4 {k}^{2}-{r}^{2}{n}^{2}-2
{\frac {m}{k\cosh \left( 2 x \right) r}}&k\cosh \left( 2 x \right)
&2 k\sinh
 \left( 2 x \right) r&-n{r}^{2}\\\noalign{\medskip}k\cosh \left( 2 x
 \right) &0&0&0\\\noalign{\medskip}2 k\sinh \left( 2 x \right) r&0&-
{r}^{2}&0\\\noalign{\medskip}-n{r}^{2}&0&0&-{r}^{2}\end {array}
 \right),
\end{eqnarray}
where $k,~m,~n$ are constants.

\section{Conclusion}
\setcounter{equation}{0}

All the examples mentioned above always contain some black holes,
i.e. singularities in Riemann tensor $R_{\mu\nu\al\be}$,  so they
are not trivial solutions. These solutions show that the canonical
metric is really convenient and powerful for solving the vacuum
Einstein equation, and the procedure of resolution is quite
straightforward.

There are many exact solutions collected in the books
like\cite{mac,haw}. Most of them can be transformed into the
canonical form (\ref{1.01}). Some authors have realized the
convenience of this kind of form related to the light cone, and
the Robinson-Trautman type solutions are simple cases of
(\ref{1.01}). However these authors seem to pay attention mainly
on using the advanced mathematical tools to certain special cases.

The main characters of our procedure is that:

(I) It presents a more general framework for getting exact
solutions to the Einstein equation, which covers most of the known
results for empty space.

(II) The procedure of resolution shows that the Einstein equation
of (\ref{1.01}) is an underdetermined system, so that we should
take the components $a,~b$ as given functions determined by
coordinate conditions and boundary or initial conditions.

(III) When we take $a,~b$ as given functions, the other components
are decoupling variables in the system of equations in some sense,
so we can use the recursive method to solve the Einstein equation.

(IV) The ansatz (\ref{2.4}) or (\ref{2.23}) is natural and
necessary for getting exact solutions. As to the numerical
solution, instead of this ansatz, we should input the appropriate
values of $a,~b$ as boundary or initial conditions to determine
other components.

\section*{Acknowledgments}

The author is grateful to his supervisor Prof. Ta-Tsien Li for his
encouragement and guidance. Thanks to Prof. S. Klainerman for
suggesting me how to modify my original paper. Prof. Hao Wang
helped me to effectively use the softwares.

\end{document}